\documentclass[a4,twocolumn,amsmath,amssymb,nofootinbib,eqsecnum,floatfix,pre]{revtex4} 
\usepackage{epsfig}
\psfigdriver{dvips}
\usepackage{graphicx}
\usepackage{color}
\usepackage{dcolumn} 
\usepackage{bm, bbm}      
\usepackage{curves}
\usepackage{epic}
\usepackage{wasysym}
\usepackage{subfigure}
\def\beq{\begin{equation}}
\def\eeq{\end{equation}}
\def\bea{\begin{eqnarray}}
\def\eea{\end{eqnarray}}
\def\nn{\nonumber}

\begin{document}
\title{Mixed Columnar-Plaquette Crystal of correlated fermions on the 2D
  pyrochlore lattice at fractional filling}
\date{\today}

\author{F.~Trousselet}
\affiliation{Laboratoire de Physique Th\'{e}orique, CNRS, Universit\'{e} Paul
  Sabatier, 31062 Toulouse, France.} 

\author{D.~Poilblanc}
\affiliation{Laboratoire de Physique Th\'{e}orique, CNRS, Universit\'{e} Paul
  Sabatier, 31062 Toulouse, France.} 

\author{R.~Moessner}
\affiliation{Max-Planck-Institut f\"{u}r Physik Komplexer Systeme, 01187
  Dresden, Germany.} 

\begin{abstract}
We study a model of strongly correlated $S=1/2$ fermions on the planar
pyrochlore, or checkerboard, lattice, at fractional ($1/8$) filling. Starting
with the extended Hubbard model in the limit of strong Coulomb repulsion,
low-energy configurations can be mapped onto hard-core dimer configurations
whose dimers carry a spin degree of freedom. An effective Hamiltonian
similar to the kinetic term of a quantum dimer model on the square lattice
which rotates two parallel  dimers (in a spin-singlet configuration) by
90$^o$-degrees naturally emerges. We also introduce an additional term in the
Hamiltonian, a generalized dimer plaquette interaction, in order to realize a
closer analogy to the latter model. For a strong dimer plaquette
attraction stabilizing a columnar phase, a
spontaneous dimerization takes place in the direction of the columns of
(spin-carrying) dimers. Using exact diagonalizations of two-dimensional
periodic clusters, the analysis of the low-energy spectrum and of several types
of correlation functions gives indeed evidence for a new type of lattice symmetry
breaking phase, the eight-fold degenerate Mixed Columnar-Plaquette Crystal, and
for a transition from this phase to a Resonating-Singlet-Pair Crystal (found in
previous studies) which restores the rotational symmetry of the
lattice. Similar conclusions and phase diagram are also reached from a simple
variational approach. 
\end{abstract}

\maketitle
\section{Model, purposes, and method}
\subsection{Introduction and summary of previous results}

\begin{figure}
\begin{center}
\includegraphics[width=4cm]{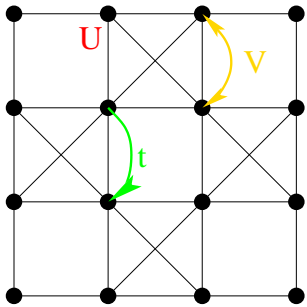}
\caption{\label{diag1} Schematic representation of the Hubbard model on the
  checkerboard lattice. In the limit $|t| \ll V \ll U$, this model becomes
  equivalent to that described by the Hamiltonian $H_{K}$ 
(see Eq.~\protect\ref{hk})}
\end{center}
\end{figure}

The interplay between electronic correlations and the lattice geometry in
quantum magnets can lead to a rich variety of spin gapped disordered phases,
either spin liquids with fractionalized excitations or various types of Valence
Bond Crystals (VBC), which break spontaneously some of the lattice symmetries.
Among materials magnetically frustrated and possibly presenting such phases,
those with a pyrochlore structure, a three-dimensional (3D) array of
corner-sharing tetrahedra, are of particular interest because of the absence of
magnetic order down to very low temperatures~\cite{pyro}. On a two-dimensional
(2D) version of the pyrochlore lattice, the \textit{checkerboard} lattice (see
Fig.~\ref{diag1}), the Heisenberg model presents a VBC of particular interest,
the \textit{plaquette} phase~\cite{placH}, which exhibits the rotational
symmetry of the lattice. To understand the physics of undoped and doped
frustrated magnets and predict the occurrence of these phases in real
materials, 
theoretical tools such as the Hubbard model, and models derived from it in the
limit of strong on-site repulsion, are commonly used in 2D (also in 3D)
systems. These exhibit very interesting properties: in a model of bosons on the
triangular lattice, doping away from commensurate fillings drives a transition
from an insulator to a supersolid phase (with charge ordering and a finite
superfluid density) ~\cite{3tri}, which is also found in a model of bosons on
the checkerboard lattice~\cite{Sen2}; on the same lattice, spinless fermions
near $1/4$ filling present interesting properties such as fractional charge
excitations ~\cite{PBSF,FPS}. 

In a more specific context, to describe the non-magnetic Resonating
Valence-Bond phase of cuprate materials and the transition of this phase to the
superconducting phase, the Quantum Dimer Model (QDM) was developed in the late
eighties~\cite{RK}, mostly on two-dimensional lattices. This model displays
different types of Valence Bond Crystals (with close analogies with their spin
counterparts), among which the quite exotic \textit{plaquette} phases, and,
depending on the lattice (non-) bipartiteness, either a liquid phase with
topological order (on the triangular lattice~\cite{MS03}) or a quantum critical
point (Roksar-Kivelson point), both presenting deconfined excitations \cite{MS03}. The QDM
is also connected to the physics of pyrochlore systems, since strong Coulomb
repulsion in the (extended) Hubbard model on the kagom\'{e} or on the
checkerboard lattice, either for bosons or fermions at special fractional
fillings, select low-energy configurations that can be mapped onto dimer or
loop configurations. 

The present study belongs to a series of works about the extended Hubbard model
on the 2D pyrochlore (or checkerboard) lattice, at and slightly away from
fractional fillings~\cite{PPS,tJ07,PBSF,WesB}. Here we focus on the effective 
model of $S=1/2$ fermions at 1/8 filling (1 particle for 4 sites)  derived from the
extended Hubbard model, 
\begin{eqnarray}
H= -t \sum_{<ij>,\sigma} c^{\dagger}_{i,\sigma}c^{\dagger}_{j,\sigma}
+U\sum_{i} n_{i\uparrow} n_{i\downarrow} +V\sum_{<i,j>} n_{i}n_{j}\, , 
\end{eqnarray}
in the limit of large Coulomb repulsions $U$ and $V$ (with respect to the
energy scale given by the particle hopping amplitude $t$). Let us briefly
review our current understanding of the physics of the extended Hubbard model on 
the pyrochlore lattice at the special fractional fillings $n=k/4$
$(k=1,2,3)$~\cite{PPS,tJ07} and mention the important remaining issues. 
For spinful fermions at these fillings, in the limit where the on-site repulsion
$U$ is very large compared to the nearest neighbour repulsion $V$ and hopping
$t$, a metal-to-insulator transition was found for increasing $|V/t|$ at the 
filling factor $n=1/4$ (similarly as in the corresponding hard-core bosonic 
model~\cite{WesB}), and the corresponding insulating phase (for $|t| \ll V \ll U$)
exhibits plaquette correlations indicating an ordering very different from a
simple charge density wave. Whether the metal-to-insulator transition occurs
immediately at infinitesimal $V$ or at a finite value depends on the degree of 
the commensurability $k$. Indeed, the perfect nesting
property of the non-interaction Fermi surface realized only for $k=2$ (and for
a given sign of $t$) leads to an instability for arbitrarily small $U$ and
$V$~\cite{Indergand}. In the limit of interest here (strong couplings), an
effective model of $S=1/2$ fermions was derived in the same study, involving a
2-particle hopping term (amplitude $t_{2}$) and an additional term (amplitude
$W$) counting the number of singlet pairs on uncrossed plaquettes. Varying the
ratio of the amplitudes of these terms, the system can be tuned from a charge
ordered \textit{columnar} phase (the internal structure of columns being that
of Heisenberg AF chains) in the limit $W \ll -|t_{2}|$ to a disordered RK point
at $W=t_{2}$. Note that, in this formulation, the case $W=0$ is believed to
provide the effective description of the insulating phase of the large-U,
large-V Hubbard model on the checkerboard lattice mentioned above. 

So far the case corresponding to filling $n=1/2$ is understood the best. A
phase transition was clearly evidenced at finite (negative) $W/t_{2}$ between
the charge-ordered phase and a Resonating Singlet-Pair Crystal (RSPC) using an
analysis based on the symmetry-resolved low-energy spectrum and plaquette
correlations in the ground state (GS). The system at $n = 3/4$ was shown to
exhibit also plaquette order by forming a (lattice rotationally-invariant)
Resonating Singlet-Pair Crystal, although with a quadrupling of the lattice
unit cell (instead of a doubling for $n=1/2$) and a 4-fold degenerate ground
state. Concerning the $n=1/4$ case, the conclusions of the previous study were
less clear: although the evolution of plaquette correlations with $W/t_{2}$
also supports a transition from a charge-ordered to a RSPC, the analysis of
low-energy eigenstates was less conclusive than in the $n=1/2$ and $n=3/4$
cases, primarily due to larger finite-size effects: in the previous study
computations were done on a $N=32$ checkerboard cluster with periodic boundary
conditions. Moreover for $n=1/4$, taking into account the possibility of new
mixed phases (which are not charge-localized but break rotational symmetry)
requires more caution in the analysis of the low-energy spectrum (and hence
larger clusters). This leads us to consider a new scenario for the phase
diagram of the model, which will be described in more detail hereafter.

\subsection{\label{modeff} The effective model}

As outlined above, the effective model is derived from the extended-Hubbard
model for $S=1/2$ fermionic particles on the checkerboard lattice, in the limit
of very large Coulomb repulsion (more precisely $|t| \ll V \ll U$). In this
limit, at $1/8$ filling ($n=1/4$), one can exclude configurations where the 2
neighboring sites are simultaneously occupied. In other words, each tetrahedra
should contain exactly one particle (of either spin), an {\it ice rule}-type
constraint, which still leaves an exponentially large number of states. As
discussed in the literature, once the particles are viewed as dimers living on
the bonds of the square lattice formed by the centers of crossed plaquettes,
this constraint is equivalent to the {\it hard core} dimer constraint on
the square lattice. However, in contrast to the ``usual'' QDM on the square
lattice, here each dimer carries a color index (associated to the spin of the
electron it represents).  

In this limit, a single particle hops out of a low-energy configuration
(\textit{colored dimer} configuration) creates a defect centered on a
tetrahedra with an energy cost $V$. This defect can however be annealed by the
subsequent hopping of the second particle on the "defect tetrahedron". Such
processes lead to an effective kinetic term, i.e. a correlated 2-particle
hopping, of amplitude $t_{2}=2t^2/V$. In terms of dimers, this term looks like
the kinetic term of the RK model, but acts only on particles of opposite spin
on the same uncrossed plaquette (i.e. dimers of opposite color on the same
plaquette). The particles being fermionic, the expression of the kinetic term
involves operators of creation (destruction) of singlets on uncrossed
plaquettes,
$c^{\dagger}_{i \uparrow}c^{\dagger}_{j \downarrow}-c^{\dagger}_{i \downarrow}c^{\dagger}_{j \uparrow}$
($c_{i \uparrow}c_{j \downarrow}- c_{i \downarrow}c_{j \uparrow}$),

\begin{eqnarray}
H_{K}= -t_{2} \sum_{<ijkl>} \Bigg(
(c^{\dagger}_{i\uparrow}c^{\dagger}_{j\downarrow}-c^{\dagger}_{i\downarrow}c^{\dagger}_{j\uparrow})\times\nn\\ 
(c_{k\uparrow}c_{l\downarrow}-c_{k\downarrow}c_{l\uparrow}) +c.c. \Bigg) \, ,  
\end{eqnarray}
where the sum is on uncrossed plaquettes (going around a plaquette $<ijkl>$,
sites are in the order $i,k,j,l$). A unitary transformation, consisting in
defining operators $b^{(\dagger)}_{i\downarrow}=-c^{(\dagger)}_{i\downarrow}$ on every other
ascending and every other descending line of the checkerboard lattice oriented
as on Fig.~\ref{diag1} (i.e. every other vertical line of vertical links and
every other horizontal line of horizontal links of the associated square dimer
lattice) and $b^{(\dagger)}_{i,\sigma}=c^{(\dagger)}_{i,\sigma}$ otherwise,
allows that each 2-particle hopping term to have the same amplitude $-t_{2}$ in
terms of $b^{(\dagger)}_{i,\sigma}$ operators, 
\begin{eqnarray}
H_{K}= -t_{2} \sum_{<ijkl>} \Bigg(
(b^{\dagger}_{i\uparrow}b^{\dagger}_{j\downarrow}+b^{\dagger}_{i\downarrow}
b^{\dagger}_{j\uparrow})
\nn\times\\ 
(b_{k\uparrow}b_{l\downarrow}+b_{k\downarrow}b_{l\uparrow}) +c.c. \Bigg)\, .
\label{hk}
\end{eqnarray}
Notice that this is valid only {\it in the insulating phases} at specific
fractional fillings like $n=1/4$, thanks to the ice rule-type constraint. In
addition, it is possible to label the sites of the lattice in such a way that
all exchange processes on the empty squares do not involve any reordering of
the fermions so that the $b^{(\dagger)}_{i,\sigma}$ operators can be considered
as bosonic. In other words, our new formulation uses the bosonic representation
of the spin singlets (we have checked the equivalence numerically on the
32-site cluster). 

Following the initial suggestion of Ref.~\cite{PPS} and according to the
discussion above, we also consider a term analogous to the potential term of
the QDM, although here it is no longer diagonal in the basis of configurations,

\begin{eqnarray}
H_{W}&=&W \sum_{<ijkl>}
\Bigg(n_{i}n_{j}(1-n_{k})(1-n_{l})(1/2-2\textbf{S}_{i}.\textbf{S}_{j}) \nn \\ 
&+&n_{k}n_{l}(1-n_{i})(1-n_{j})(1/2-2\textbf{S}_{k}.\textbf{S}_{l})\Bigg) \,
\end{eqnarray}
This terms "counts" the number of singlet pairs of next-nearest neighbours 
(\textit{parallel dimers}) in all uncrossed plaquettes. The resulting Hamiltonian 
$H_{W}+H_{K}$ has a structure similar to that of the Roksar-Kivelson QDM, with 
both terms flipping dimers and terms counting the flippable pairs of dimers. 
$H_W$ can also be 
interpreted as a 4-site ring-exchange term on uncrossed plaquettes~\cite{tJ07}.
It also presents a RK 
point (here at $W=t_{2}$), while for $W/|t_{2}| \rightarrow -\infty$ an ordering 
in chains is favored; varying $W/t_{2}$ allows to make a continuous connection 
between both these limits and the case $W=0$, and to understand better the 
physics around this point.

\subsection{\label{diag} Purpose of the study: phase diagram}

Since $t_{2}$ and $W$ are the only energy scales in this model (at zero
temperature) we aim at determining the phase diagram as a function of the ratio
$W/t_{2}$ (or $W$, if we set $t_{2}=1$). First, we notice that for $t_{2}=W$,
the Hamiltonian has the same property as in the quantum dimer model at the RK
point: it can be written as a sum of projectors (one per uncrossed
plaquette). At this point (the RK point of the $t_{2}-W$ model) the wave
function with an equal amplitude on all configurations (in each sector of
connected configurations) is annihilated by each projector, and thus is a
ground state with zero energy. For $W \ge t_{2}$, again for similar reasons as
in the QDM, configurations of minimal energy are those without any flippable
pair of spins, and these configurations are (degenerate) ground states with
zero energy. The center of interest of this study is the case where $W \le
t_{2}$, i.e. the region between the RK point and the $W = -\infty$ point, where
the ground state is composed of decoupled Heisenberg chains (its energy is the
sum of the energy of these chains with an AF coupling $J=2|W|$ and a
\textit{charge} term $L_{tot}\, W/2$ where $L_{tot}$ is the cumulative length
of the chains). Instead of considering $W$ and $t_{2}$, one can define a
reduced parameter $\theta=\arctan(W/t_{2})$ varying continuously between the
decoupled Heisenberg chains ($\theta=-\pi/2$) and the RK point
($\theta=\pi/4$).  

Between these points, the different phases one can expect are  
(i) a \textit{columnar} phase, ordered in chains, with rotational symmetry 
breaking, translational
symmetry breaking in one direction (perpendicular to the chains) and thus 
a 4-fold degeneracy of the corresponding ground state; this phase is encountered 
for $\theta = -\pi/2$, and could 
a priori extend over a finite range of $\theta$ in the vicinity of that point. 
Note that the term \textit{columnar} is used although this phase differs from the 
columnar phase of the QDM on the square lattice, due to the additional spin
degrees of freedom;
(ii) a RSPC or \textit{plaquette} phase, with the full rotational symmetry of
the lattice but a breaking of translational symmetry in both directions and a
4-fold degeneracy of the GS; (iii) mixed phases, with rotational and
translational (in both directions) symmetry breaking (and a 8-fold degenerate
GS), corresponding to a dimerization of the Heisenberg chains. A priori two
types of mixed phases appear naturally depending whether neighboring chains
dimerize in phase, or in antiphase. Note that these phases are a natural
extension of the one recently discovered in the context of the RK
QDM~\cite{Arnaud}. We shall refer to them as \textit{Mixed Columnar-Plaquette
  Crystals} (MCPC). 

\begin{figure}
\begin{center}
\includegraphics[width=7.8cm]{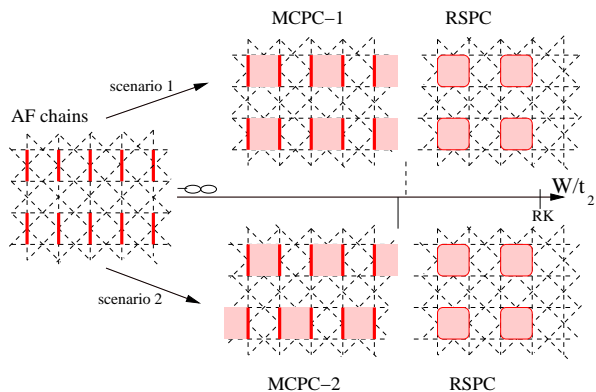}
\caption{\label{diags} (Color online) Possible scenarios for phase diagrams at 
$n=1/4$ as a function of the ratio of parameters $W/t_{2}$, depending whether an 
infinitesimal $t_{2}$ coupling dimerizes antiferromagnetic (AF) chains from a 
columnar order into a \textit{Mixed Columnar-Plaquette Crystal} 
of type 1 (\textit{MCPC-1}, up) or 2 (\textit{MCPC-2}, down). The transitions 
indicated between either of these phases and  a \textit{RSPC} should be presumably 
of first order in the first case and second order in the second case, but one
cannot exclude that the \textit{MCPC} phase extends all the way to the $RK$ point.}
\end{center}
\end{figure}

Knowing that $\theta = -\pi/2$ corresponds to a columnar phase, we refer in
phase or in antiphase dimerisation as MCPC-1 and MCPC-2, respectively, as shown
schematically in Fig.~\ref{diags}. 

These phases have distinct symmetries and the corresponding ground states are
characterized by different sets of (4 or 8) quantum numbers, which we define
using the following conventions: the $x$ and $y$ axes of the lattice are
parallel to the links of the square lattice (on which the \textit{dimers}
live), and the unit length corresponds to one link of this square lattice. To
define point group symmetries (those of the $C_{4v}$ point group - or $C_{2v}$
or $C_{v}$ for certain wave vectors) we set the center $O$ of the lattice at
the center of an uncrossed plaquette. The quantum numbers of the degenerate GS
of the various phases are listed in Table~\ref{Table}.

\begin{table}
\begin{tabular} {|c|c|c|c|c|}
\hline
Phase $\rightarrow$ & MCPC-1 & MCPC-2 & RSPC & Columnar \\
\hline
$(A1,q=(0,0))$       & X & X & X & X \\
\hline
$(B1,q=(0,0))$       & X & X & 0 & X \\
\hline
$(A1,q=(\pi,\pi))$   & X & 0 & X & 0 \\
\hline
$(A'1,q=(\pi,0))(1)$ & X & X & X & X \\
\hline
$(B1,q=(\pi,\pi))$   & X & 0 & 0 & 0 \\
\hline
$(A'1,q=(\pi,0))(2)$ & X & 0 & 0 & 0 \\
\hline
$(B',q=(\pi/2,\pi))$ & 0 & X & 0 & 0 \\
\hline
\end{tabular}
\caption{
\label{Table}
Quantum numbers of the degenerate GS (in the thermodynamic limit) associated
with the various phases expected in the $t_{2}-W$ model at $n=1/4$ on
the checkerboard lattice (for $W \le t_{2}$). The $X$ sign indicates 
that a wave function with the corresponding quantum number belongs to the 
degenerate GS manifold for the corresponding phase ($0$ otherwise).
Irreducible representations labeled with $'$ are respective to subgroups $C_{2v}$ 
$(A'1)$ and $C_{v}$ $(A',B')$ of the point group $C_{4v}$ when the wave vector 
considered is non-invariant under $C_{4v}$.} 
\end{table}

Given the symmetries of the candidate phases, one can make a guess about the
nature of phase transitions in the model, for both scenarii described in
Fig.~\ref{diags}. In the case of a \textit{MCPC-2} phase for $|t_{2}| \ll -W$,
giving way to a \textit{RSPC} close to the RK point, the transition between
those should be of first order, since the symmetry groups of one phase is not
included in that of the other. By contrast, since the \textit{MCPC-1} phase
distinguishes itself from the \textit{RSPC} by the breaking of one of its
symmetries (invariance by a $\pi/2$ rotation), the transition between those
could be of second order.   

\subsection{Methods}

In the present work, we first discuss the regime near $W/t_2=-\infty$ 
perturbatively (Section~\ref{2LH}). Next, 
implement a simple variational approach 
(discussed further in Section~\ref{vari}) adapted to describing the various
candidate phases; in a second step, we shall use Lanczos Exact Diagonalization
techniques to study the $t_{2}-W$ model on clusters with periodic boundary
conditions in both directions. The sizes of the clusters we consider in
numerics are $N=32,72$ ($\pi/4$-tilted checkerboard clusters corresponding to
untilted square clusters of lengths $L=4,6$), $N=48$ ($\pi/4$-tilted
checkerboard cluster corresponding to an untilted rectangular cluster of
dimensions $(L_{x},L_{y})=(4,6)$) and $N=64$ (untilted checkerboard cluster
corresponding to a $\pi/4$-tilted square lattice of length
$L=4\sqrt{2}$). Except for $N=72$, the cluster periodicity is compatible with
all wave vectors $\textbf{q}$ mentioned in the table above (for the $N=72$
cluster, the wave vector $\textbf {q}=(\pi/2,\pi)$-and those equivalent to it
up to point group symmetries- are unaccessible). 

Note that we restrict ourselves to the sector $S_{z}=0$ (which includes all
total spin sectors) - for convenience. In addition, we consider only
configurations for which the $z$-components of the total spin on each row (of
vertical bonds) and each column (of horizontal bonds) of the square lattice
(which are conserved quantities in the present model) are zero. This condition
is satisfied by the ground states corresponding to any of the expected phases,
and allows us to reduce the number of colorings of any dimer (=charge)
configuration (hence the total size of the Hilbert space). By using the 
 character of spin inversion, all point group symmetries and
translations (in fact, due to the numerical technique for encoding
configurations, we use translations not interchanging the sublattices of the
dimer lattice, hence $N/4$ translations instead of $N/2$) the number of
representatives for $N=64$ and $N=72$ clusters are close to $4.5 \cdot 10^4$
and $1.9 \cdot 10^5$ respectively~\cite{N72}. 

A powerful tool to determine the phase diagram is the analysis of the lowest
energy levels in symmetry sectors associated to each of the quantum numbers
mentioned (and with a character $1$ for the spin inversion $S_{i}^z \rightarrow
-S_{i}^z$). For the various phases one expects a degeneracy between the quantum
numbers marked by $X$ in the table~\ref{Table}, in the thermodynamic limit ($N
\rightarrow \infty$). On finite clusters, this degeneracy is lifted and the
lowest state is always found for $(A1,{\textbf q}=(0,0))$ - hence we look at
the lowest excited states. Ideally, an unambiguous signal of spontaneous
symmetry breaking is provided by the collapse of the corresponding  excitation
energies $\Delta E = E_{i} - E_{0}(A1,(0,0))$ with increasing $N$. However, the
low-energy spectrum on large enough a cluster (e.g. $N=64$) gives enough
information for a first analysis. 

\section{\label{2LH} Perturbative approach from $W=-\infty$: coupling of
  Heisenberg chains} 

Adding the potential term $W$, to the Hamiltonian has several benefits, among
them the existence of two particularly simple special points, namely the RK
point, and the point at $W = -\infty$. For the bosonic model, the latter
yields the simple columnar configurations as ground states. However, $W =
-\infty$ does not always present such a simple setting. For the triangular RK
model, there are two families of ground states, each comprising a number of
members exponential in the linear system size, the degeneracy between which is not
lifted until perturbations
to leading non-trivial order in $t_{2}/W$ are taken into
account \cite{MS03}.

The situation for our model is different still. {\it At} $W = -\infty$, the
ground state is obtained by maximising the number of plaquettes with a pair of 
dimers in a singlet configuration. This leads to formation of a state breaking
rotational and translational symmetries --- just like the columnar state ---
but in which the spin correlations along the columns are critical. Indeed, $W =
-\infty$ corresponds to decoupled Heisenberg cains.

The question appropriate to the setting of small $|t_{2}/W|$ is thus: what is the
most relevant perturbation induced by the kinetic term. This question has been
addressed in --- formally related --- contexts by Essler, Tsvelik and Starykh +
coworkers \cite{Starykh, Essler}. 
We closely follow the approach of the latter \cite{Starykh}. Their
observation that in the Heisenberg chain not only the staggered spin but also
the staggered energy correlations are critical --- both decay as $1/r$ --- is
central: the chains are close to not only Neel but also to dimer ordering.

\begin{figure}
\begin{center}
\includegraphics[width=6cm]{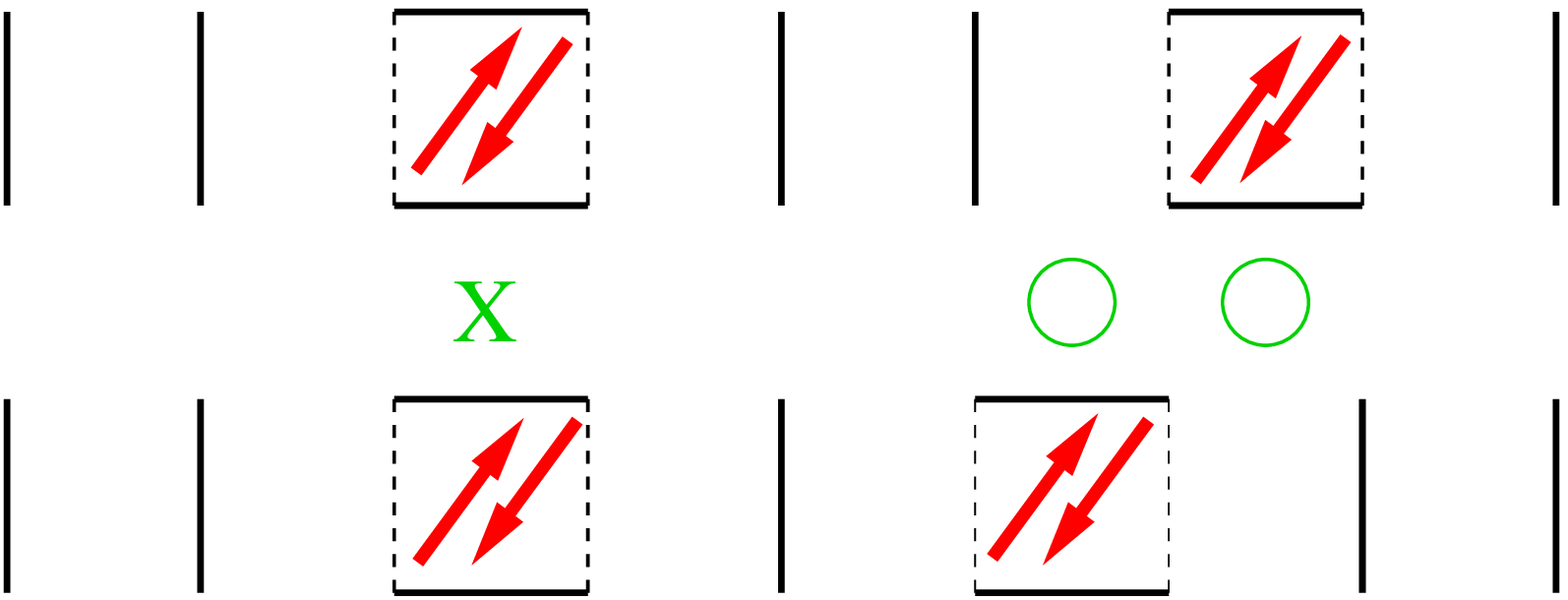}
\caption{\label{PM} (Color online) Origin of dimerisation in the limit of large 
negative $W$ (singlets flipped by $H_{K}$ are represented by an up-down spin pair):
the plaquette marked by a cross gains an energy $W/2$, 
while there is no such gain on plaquettes marked by circles.}. 
\end{center}
\end{figure}

For our model, the coupling of the
staggered dimerisation between neighbouring chains is symmetry
allowed, and hence will generically appear as a perturbation is
added. Indeed, it is easy to see how this happens in our model. For
finite $|t_{2}/W|$, flipping the dimers in two neighbouring plaquettes in 
adjacent rows yields an energy gain of $O(W)$ for the plaquette marked by a cross in Fig.~\ref{PM}, whereas there is no such gain for the two plaquettes
marked by circles. the coupling between the chains is thus generated
at $O(|t_{2}/W|^4)$, as each plaquette needs to be flipped out of the
chain and back.

As analysed in Ref.~\cite{Starykh}, this coupling is relevant and it will
immediately lead to an in-phase dimerisation of adjacent chains. In our above
classification, this corresponds to a  \textit{MCPC-1} phase, which is thus
present in the limit $W/t_{2} \rightarrow -\infty$.

Before we move on to variational (Section \ref{vari}) and numerical
(Sections \ref{exd} and \ref{corr}) investigations of the effective
$t_{2}-W$ model on the checkerboard lattice, let us make some further
remarks on the regime where $W/t_{2} \ll -1$. This will serve as a
warm-up exercise for identifying ground states by quantum numbers. Let
us consider a pair of Heisenberg chains of length $L$ (with periodic
boundary conditions, the chain axis being $x$), of spins-1/2 $S_{i,1}$
and $S_{i,2}$. The Hamiltonian,
\begin{eqnarray*}
H=J \sum_{i,\alpha} \textbf{S}_{i,\alpha}.\textbf{S}_{i+1,\alpha} + K
\sum_{i}(\textbf{S}_{i,1}.\textbf{S}_{i+1,1})(\textbf{S}_{i,2}.\textbf{S}_{i+1,2})\,~~.   
\end{eqnarray*}
captures then coupling described above, although the true effective coupling is
more complicated, in particular, it includes similar four-spin terms along the
chain favouring the same dimerisation pattern. 

At $K=0$, the ground state is the product of ground states of the Heisenberg
model on each chain, and has a symmetry $(A,k_{x}=0)$-- A (B) labeling even
(odd) states w.r.t. the chain interchange. The first excited state in the $S=0$
sector correspond to a 2-triplet excitation on one chain (the other chain
remaining in the GS). It is doubly degenerate, the quantum numbers of both
states being $(A,k_{x}=\pi)$ and $(B,k_{x}=\pi)$. This state has (similarly to
1-triplet excitations) an excitation energy proportional to $1/L$, thus
collapsing to the ground state in the thermodynamic limit. 

\begin{figure}[!h]
\begin{center}
\includegraphics[width=8cm]{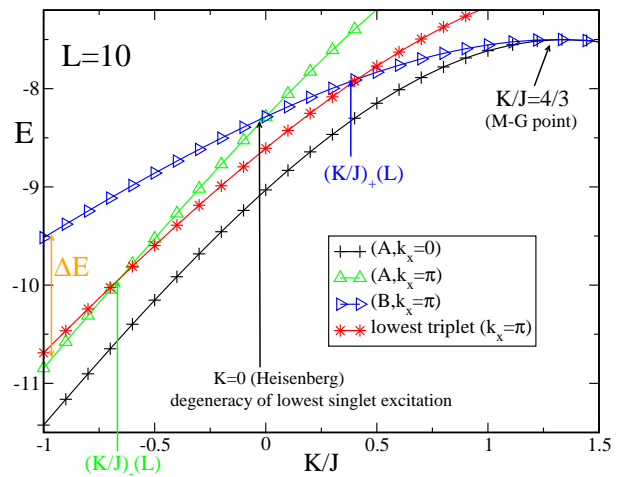}
\caption{\label{JK10} (Color online) Energies of the ground state and first 
excited states of two periodic Heisenberg chains of $L=10$ spins 1/2 (Heisenberg 
coupling $J$) coupled with a 4-spin interchain coupling $K$. $(K/J)_{c}$ indicates
 the crossing between the lowest triplet and the $(A,k_{x}=\pi)$ singlet. Notice
  also the degeneracy of $(B,k_{x}=\pi)$ and $(A,k_{x}=0)$ states at the
  Majumdar-Ghosh point $K/J=4/3$.} 
\end{center}
\end{figure}

When a weak interchain coupling $K$ is added, the degeneracy of this first
excited state is lifted, with a splitting proportional to $K/J$, as shown on
Fig.~\ref{JK10}. Since each of these states has the symmetry of a dimerized
state, with dimerization either in phase (state $(A,\pi)$) or in antiphase
(state $(B,\pi)$), the sign of this splitting is therefore associated with the
type of dimerization susceptible to spontaneously appear in the system: for
$K/J \ge 0$, the $(B,\pi)$ state has a lower energy than the $(A,\pi)$ state
and the system tends to dimerize in antiphase; this is the opposite case for
$K/J \le 0$. 

\begin{figure}
\begin{center}
\includegraphics[width=7cm]{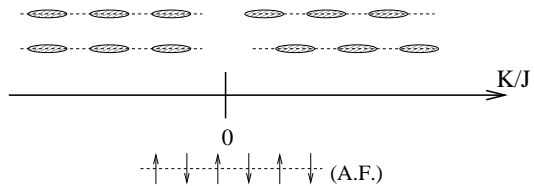}
\caption{\label{JK} Types of dimerization occurring in 2 Heisenberg chains
  coupled by a 4-spin coupling: in-phase dimerization ($K \le 0$) or antiphase
  dimerization ($K \ge 0$)}. 
\end{center}
\end{figure}

\begin{figure}[!h]
\begin{center}
\includegraphics[width=8cm]{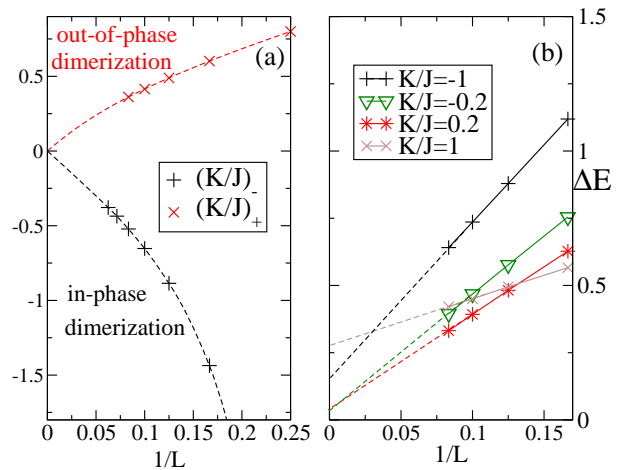}
\caption{\label{Kc} (a) (Color online) Position of the energy crossings between 
the lowest triplet and lowest $(A,\pi)$ ($(K/J)_{-}$) and $(B,\pi)$ ($(K/J)_{+}$)
  singlet states as a function of $1/L$ ($L$ even up to $16$); (b) Excitation
  energy $\Delta E$ of the lowest triplet for different values of $K/J$  versus
  $1/L$. Dashed lines are only a guide to the eyes.}
\end{center}
\end{figure}

Let us examine now the finite size scaling of these excitations in
presence of the 4-spin coupling. The excitation energy of the lowest
excited singlet state ($(A,\pi)$ for $K \le 0$ and $(B,\pi)$ for $K
\ge 0$) vanishes in the thermodynamic limit (the convergence is as
$1/L$ for $K/J \ll 1$), while the excitation energy to the lowest
triplet converges to a finite value (spin gap) (see right plot on
Fig.~\ref{Kc}).  Consequently, while on a small enough system the
triplet excitation has lower energy than the lowest singlet, these
levels cross as a function of system size (at fixed $K/J$). This is
illustrated (in a slightly different way) by the left plot on
Fig.~\ref{Kc}, showing the positions of crossings $(K/J)_{+/-}(L)$
between the lowest triplet and either the lowest $(A,\pi)$ or
$(B,\pi)$ singlet, as a function of system size. One can see that
these values converge to zero as $L \rightarrow \infty$. (Rigorously,
one cannot be fully conclusive with the present data set but a
vanishing value as $L \rightarrow \infty$ is expected from the
reasoning above. Moreover we checked that expressions of the type
$C/L+C'/L^2$ fit the data of $(K/J)_{+/-}$ better than any expression
of the type $C+C'/L^\alpha$.)  Therefore the lowest singlet
excitation, of symmetry either $(A,\pi)$ or $(B,\pi)$ depending of the
sign of $K$, collapses onto the GS in the thermodynamic limit while a
spin gap survives above. This is the well-known scenario of a
spontaneous dimerization, which is precisely the type of scenario one
expects in our two-dimensional effective model in the limit of weakly
coupled chains (of colored dimers). Here, it is important to note that
there are distinct reasons for the vanishing gaps. Even for $W =
-\infty$, there will be finite-size gaps of $O(|W|)$, which vanishes
algebraically due to the criticality of the chains, while the other gap, 
parametrically small in $|t_{2}/W|$, collapses due to the
presence of symmetry-breaking.

\parskip 5pt

\section{\label{vari} Variational approach describing the candidate phases}
\subsection{Principle and trial wave functions}

Before analyzing the exact ground state and lowest excitations on finite
systems, we estimate the energies of trial wave functions associated with the
different candidate (\textit{RSPC}, \textit{MCPC-1} and \textit{MCPC-2}) phases
in the range of parameters of interest ($-\pi/2 \le \theta \le \pi/4$). A
comparison between their variational energies provides information on their
relative stability. The trial wave functions we consider are built as the
tensor product of all equivalent plaquettes of an identical wave function
$|\psi_{p}\rangle$ defined on a single uncrossed plaquette $p$, with a
resonating singlet delocalized on the 4 sites of the plaquette. 

On this plaquette, $|\psi_{p}\rangle$ is expanded over the 4 $S_{z}=0$
configurations with 2 particles on the plaquette (and respecting the
\textit{dimer} constraint): if the 4 sites of the plaquette are labeled from 1
to 4 clockwise around the plaquette (starting from e.g. the site on the upper
left side) these 4 configurations are: $|u\rangle=|\uparrow_{1} 0_{2}
\downarrow_{3} 0_{4} \rangle$ ($0$ denoting an empty site);
$|d\rangle=|\downarrow_{1} 0_{2} \uparrow_{3} 0_{4} \rangle$; $|r\rangle=|0_{1}
\uparrow_{2} 0_{3} \downarrow_{4} \rangle$ and $|l\rangle=|0_{1} \downarrow_{2}
0_{3} \uparrow_{4} \rangle$. 
$\psi_{p}(\phi)$ is a linear combination of the 2 singlet states
$(|u\rangle +|d\rangle)/\sqrt{2}$ and $(|r\rangle+|l\rangle)/\sqrt{2}$ (the
plus sign results from the unitary transformation mentioned in \ref{modeff}):\
$$|\psi_{p}\rangle(\phi)=\cos(\phi)\frac{|u \rangle +|d\rangle
}{\sqrt{2}}+\sin(\phi)\frac{|r\rangle + |l\rangle }{\sqrt{2}}$$. 
The parameter $\phi$ can be restricted to values between $0$ and $\pi/4$, thus
describing mixed phases obtained by dimerization of horizontally oriented
Heisenberg chains. The global wave function on a $N$-site cluster
$|\Psi_{0}(\phi)\rangle=\otimes^{N/8}_{i=1} |\psi_{p_{i}}(\phi)\rangle$
depends on the angle $\phi$ and the set of $N/8$ chosen plaquettes ${p_{i}}$,
i.e. the type of dimerization described: either in phase (\textit{MCPC-1}) or 
in antiphase (\textit{MCPC-2}). In the first case the \textit{RSPC} corresponds
to an angle $\phi=\pi/4$. The wave function $|\Psi_{0}\rangle$ need not be
symmetrized w.r.t. space group symmetries, in order to compute of
the expectation values of $H_{K}$ and $H_{W}$ in the thermodynamic limit: 
indeed, for two symmetry-related and distinct vectors $|\Psi_{0}\rangle$ 
and $|\Psi'_{0}\rangle$, quantities such as $\langle \Psi'_{0}|H_{K/W}|\Psi_{0}
\rangle$ give a relative contribution $O(1/N)$.  

\subsection{Trial function for \textit{MCPC-1} and \textit{RSPC} phases}

In order to describe the \textit{MCPC-1} phase and its stability w.r.t. the
\textit{RSPC}, the trial wave function $|\Psi_{0}^{1}\rangle(\phi)$ is used to
compute expectation values of $H_{K}$ and $H_{W}$. The only terms of $H_{K}$
contributing to $\langle \Psi_{0}^{1}|H_{K}|\Psi_{0}^{1} \rangle$ are plaquette
flips on occupied plaquettes, exchanging \textit{u/d} configurations and
\textit{r/l} ones. The average kinetic energy (per particle) is then:
\begin{eqnarray}
H_{K,1}(\phi)=4/N \langle \Psi_{0}^{1}|H_{K}|\Psi_{0}^{1} \rangle \nn\\
= -2 t_{2} \sin(\phi)\cos(\phi)
\label{Ht2}
\end{eqnarray}
For the non-diagonal part of $H_{W}$ (with $S_{i}^{+}S_{j}^{-}$ terms), again
only terms for which both sites are in an occupied plaquette contribute. But
concerning the diagonal part of $H_{W}$, terms $1/2-2S_{i}^{z}S_{j}^{z}$ with
$i,j$ on a void plaquette between 2 occupied plaquettes also contribute,
proportionally to either $\cos(\phi)^4$ or $\sin(\phi)^4$ depending on the
position of this void plaquette. The expectation value of $H_{W}$ (still per
particle) is: 
\begin{eqnarray}
H_{W,1}(\phi)= 4/N \langle \Psi_{0}^{1}|H_{W}|\Psi_{0}^{1} \rangle \nn\\
= W (1 + \sin(\phi)^{4}/4+ \cos(\phi)^{4}/4)
\label{HW1}
\end{eqnarray}
The minimization of $\langle H_{K}(\phi) \rangle + \langle H_{W}(\phi) \rangle$
w.r.t. $\phi$ can lead to 2 distinct cases: 

For $W/t_{2} \le -4$ a non-trivial value $\phi_{1}$ of $\phi$ minimizing the
expectation value of $H=H_{K}+H_{W}$ is found, corresponding to a
\textit{MCPC-1} state breaking the $\pi/2$-rotational symmetry. $\phi_{1}$ is
solution of: $W/t_{2}=-4/\sin(\phi_{1})$, and the corresponding average energy
is:  
\begin{eqnarray*}
E_{1}(\phi_{1}) = 2\frac{t_{2}^2}{W} + 5W/4
\end{eqnarray*}

\begin{figure}[!h]
\begin{center}
\includegraphics[width=8cm]{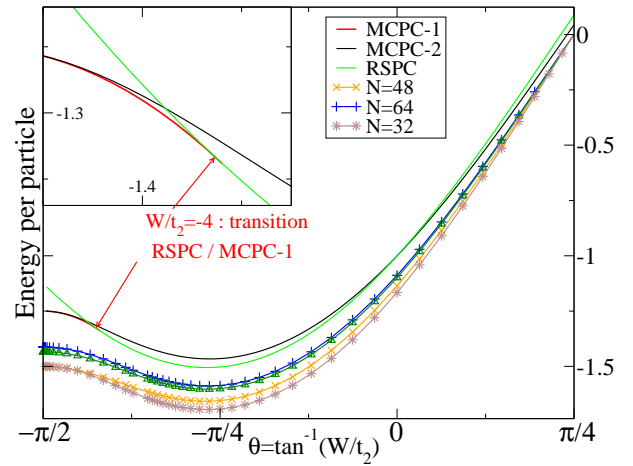}
\caption{\label{var} (Color online) Energies (per particle and in units of
  $\sqrt{t_{2}^2+W^2}$) of variational ground states describing the
  \textit{MCPC-1} [$\theta \le \arctan(-4)$], \textit{RSPC} and \textit{MCPC-2}
  phases as a function of $\theta$. Exact ground state energies for $N=32, 48,
  64$ and $72$ clusters are shown for comparison.}
\end{center}
\end{figure}

For $W/t_{2} \ge -4$ the minimization gives $\phi_{1}=\pi/4$, which means that
the rotationally invariant \textit{RSPC} is the most favorable state in this
approach. The average energy of the \textit{RSPC} is estimated as:
\begin{eqnarray*}
E_{1}(\pi/4) = -t_{2} + 9W/8
\end{eqnarray*}
The energies $E_{1}(\pi/4)$ and (for $\theta \le \arctan(-4)$)
$E_{1}(\phi_{1})$ are shown as a function of the parameter $\theta$ in figure
\ref{var}. Notice that for $\theta \le \arctan(-4)$, $H_{1}(\pi/4)$ corresponds
to a local maximum of the function $H_{1}(\phi)$ and is shown only for
comparison to the (physically relevant) \textit{MCPC-1} variational energy
$E_{1}(\phi_{1})$. 

These results give also an indication of the nature of the \textit{MCPC-1/RSPC}
transition: the expansion of the variational energy functional
$H_{K,1}+H_{W,1}(\phi)$ in powers of $x=\phi-\pi/4$, for $W/t_{2}$ close to
$-4$, is: $$H_{1}(\phi)=H_{1}(\pi/4)+2(t_{2}+W/4) x^2 + (-19W/24 -2t_{2}/3)
x^4$$ The sign of the coefficient of the $x^2$ term changes for $W=-4t_{2}$
while that of the $x^4$ term remains positive around that point: in the frame
of Landau's theory of phase transitions, this is characteristic of a second
order transition with $x$ as an order parameter, which varies
continuously around the \textit{MCPC-1/RSPC} transition and vanishes in the
$\pi/2$-rotationally invariant phase. 

\subsection{Trial function for the \textit{MCPC-2} phase}

In this case, the trial wave function $|\Psi_{0}^{2} \rangle$ is still
parameterized by an angle $\phi$ between $0$ and $\pi/4$ and differs only from
the wave function describing the \textit{MCPC-1} phase by the position of the
occupied plaquettes. The angle $\pi/4$ corresponds here to a rotationally
non-invariant pattern of rotationally invariant occupied plaquettes (which
differs from the \textit{RSPC}). Again, the structure of the trial state
implies that the contribution to the expectation values of $H_{K}$ and the
non-diagonal part of $H_{W}$ comes only from terms acting independently on
occupied plaquettes. But here the diagonal part of $H_{W}$ acts differently on
the wave function than in the previous case: it gives a non-zero expectation
value only on the void plaquettes situated between 2 occupied plaquettes to the
right and left (see Fig.~\ref{diags}). The corresponding term in $\langle H_{W}
\rangle$ is thus proportional to the probability $\cos(\phi)^4$ for particles
of both plaquettes to be in $|r\rangle$ or $|l\rangle$ states. Eventually the
expectation value of $H$ as a function of $\phi$ reads:
\begin{eqnarray}
H_{2}(\phi)= 4/N \langle \Psi_{0}^{2}|H_{W}+H_{K}|\Psi_{0}^{2} \rangle \nn\\
= W (1 + \cos(\phi)^4 /4) -2t_{2}\cos(\phi)\sin(\phi)
\label{HW2}
\end{eqnarray}
and is minimized, either for $\phi_{2}=\pi/4$ when $W \ge 0$, or when $W \le 0$
for $\phi_{2}$ solution of:
\begin{eqnarray*}
W/t_{2}=-4\frac{1+\tan(\phi_{2})^2}{\tan(2\phi_{2})}
\end{eqnarray*}
The angle $\phi_{2}$ (for $W \le 0$), and consequently the corresponding 
expectation value $E_{2}(\phi_{2})$ of $H$ ,
  has no simple expression as a function of $W/t_{2}$ or $\theta$; a numerical
  resolution proves that for $W \le 0$ it is greater than $H_{1}(\phi_{1})$
  found with the wave function $|\Psi_{0}^{1}\rangle $ (see Fig.~\ref{var}) -
  this is not a surprise since the $(W/4)\sin(\phi)^{4}$ term present in
  Eq.~\ref{HW1} is absent in Eq.~\ref{HW2}. In other words, this approach
  indicates that the \textit{MCPC-1} or \textit{RSPC} phase is stabilized
  w.r.t. the \textit{MCPC-2} phase by interactions on some plaquettes, as soon
  as they are attractive ($W < 0$). 

For $W \ge 0$ the variational ground state $|\Psi_{0}^2\rangle$ corresponding
to an angle $\phi_{2}=\pi/4$ has an energy $17W/16-t_{2}$ lower than that of
the RSPC found before. Hence this can be considered as the variational ground
state in this approach, predicting a domain of stability $-4t_{2} < W < 0$
for the \textit{MCPC-1} phase (see Fig.~\ref{2diags}) - but one has to take
into account the limitations of this approach, discussed in \ref{fiab}. 

\subsection{Comparison to the bosonic case}

In this paragraph we apply the previous variational method to the bosonic case,
i.e. to the QDM of Roksar and Kivelson on the square lattice, to have a
comparison between the variational and exact phase diagrams. In this case, a
variational wave function $|\Psi\rangle(\phi)$ describing the plaquette phase
and the mixed phase (both described in \cite{Arnaud}, i.e. the bosonic analog
of the \textit{MCPC-1} phase ; similarly a \textit{mixed 2} phase can be
defined as the bosonic analog of the \textit{MCPC-2} phase) is defined, as
previously, as a product of local wave functions on plaquettes occupied in a
plaquette pattern: 
$$|\Psi_{p}\rangle =\cos(\phi) |v\rangle + \sin(\phi)
|h\rangle$$ ($|v\rangle$ and $|h\rangle$ correspond to either 2 vertical or 2
horizontal dimers on the given plaquette). The energy per particle of the state
$|\Psi\rangle(\phi)$ is here ($-t$ and $V$ being the amplitudes of the kinetic
and potential terms in the QDM, defined as in \cite{Arnaud}):
$$H_{1}(\phi)=-2t\sin(\phi)\cos(\phi)+ V (1 + \cos^4(\phi) + \sin^4(\phi) )$$
For $V \ge -t$ this is minimized for $\phi=\pi/4$ which corresponds to a
plaquette state, while for $V \le -t$ the angle $\phi_{1}$ is such that
$V/t=-1/sin(2\phi_{1})$. The corresponding energy per particle is:
\begin{eqnarray*}
E_{1}(\phi_{1})=\frac{t^2}{4V}+V \hspace{20pt}   (V/t \le -1) \nn\\
E_{1}(\pi/4)=3V/4-t/2  \hspace{20pt}  (V/t \ge -1)
\end{eqnarray*}
As in the fermionic case, one can define a variational wave function
parametered by an angle $\phi'$ and corresponding to the \textit{mixed 2}
phase, and show that, after minimization w.r.t $\phi'$ it has a lower energy
than $E_{1}(\pi/4)$; but again, in that case the variational method is not
adapted to the situation close to the RK point. 

\begin{figure}[!h]
\begin{center}
\includegraphics[width=8cm]{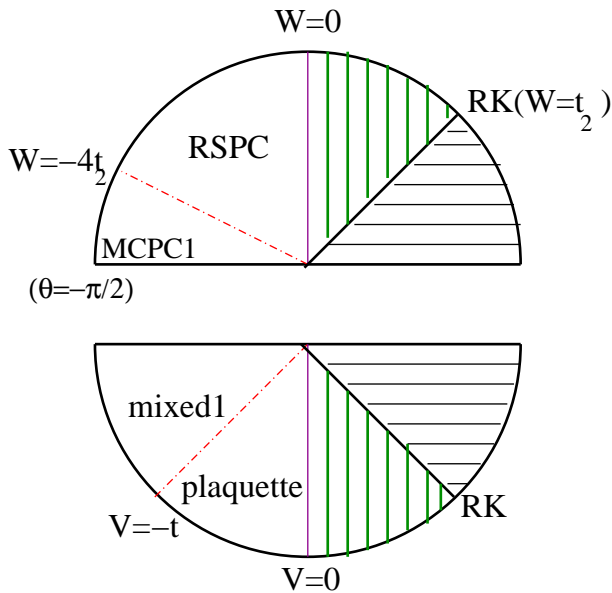}
\caption{\label{2diags} (Color online) Variational phase diagrams for the model 
of fermions (up) or bosons (down, with notations of ref.~\cite{Arnaud}).
The point of transition between a $\pi/2$-rotational invariant phase and a
  phase breaking that symmetry is predicted at $W=-t_{2}$ (for fermions) and
  $V=-t$ (for bosons). For $0 \le W \le t_{2}$ (resp. $0 \le V \le t$) the
  \textit{MCPC-2} or \textit{mixed-2} phase is the most competitive of our
crystalline trial wavefunctions, but it in turn loses out to the 
simple RK wave function, see \ref{fiab} (green vertical lines).} 
\end{center}
\end{figure}

Consequently, the variational approach for bosons on the checkerboard at
$n=1/4$ predicts the existence of a \textit{mixed} phase for $V \le -t$ and a
transition (second order again) to a plaquette phase at $V=-t$. In this
approach the plaquette phase extends from that point up to $V=0$, while the
ground state energy in the domain $0 \le V \le t$ is better approximated with a
\textit{mixed 2}-type variational wave function- this is essentially due to the
inadequacy of this method close to the RK point. The variational phase diagrams
for both the fermionic and the bosonic model (both for an average occupation
number $n=1/4$) are shown on Fig.~\ref{2diags}.  

The larger extent of the \textit{RSPC} phase in the fermionic model than of the
plaquette phase in the bosonic can be attributed to the fact that, in the first
case, a singlet resonating on a plaquette in a \textit{RSPC} phase is less
coupled to neighbouring occupied plaquettes: the process coupling neighbouring
plaquettes is subject to the constraint that interacting particles have
opposite spin, this constraint being absent in the bosonic case. Consequently,
for fermions the \textit{RSPC} phase is more stable and the transition to a
\textit{MCPC} phase with longer range correlations occurs for a larger
(negative) value of $W$. 

\subsection{\label{fiab} Reliability of the variational approach - comparison 
to exact ground state energies}

Let us now comment briefly on the reliability of this variational approach to
give a qualitative, or even quantitative, estimation of the phase diagram. For
this, we have compared the variational ground state energies $H_{1}(\phi_{1})$
and $H_{2}(\phi_{2})$ to the exact ground state energies obtained on periodic
clusters of size $N=32, 48, 64$ and $72$ (see Fig.~\ref{var}). Since the
orientation and geometrical shape of these clusters differ from each other, one
cannot do an accurate finite size scaling of the exact ground state energy that
would allow for a precise comparison between exact and variational energies in
the thermodynamic limit. However, it appears clearly that finite-size effects
on the exact energies (per particle) decrease with $N$, which suggests that for
a wide range of $\theta$ (between $-3\pi/8$ and $0$) the exact and variational
ground state energies differ from about $5\%$ or less, and their variations
with $\theta$ are very similar \cite{W0}. The discrepancy between exact and
variational results is larger (\textit{i}) when $\theta$ gets close to $\pi/4$:
at the RK point, the ground state energy on any cluster considered is zero
while the variational energy per spin is $t_{2}/16$ and corresponds to a
\textit{MCPC-2} phase unexpected here: the failure of the variational
approach close to the RK point originates in the inadequacy of the
variational wave function when the exact ground state is much more disordered 
(which holds also in the bosonic case); there, the RK wave function is a
better trial wave function, with an energy per particle $E_{RK}=W-t_{2}$
[$(V-t)/2$ in the bosonic case] \cite{RKvar}. This RK trial wave function has
an energy lower than $|\Psi_{0}^2\rangle$ for either $W/t_{2}\ge 0$ or $V/t \ge
0$ (see Fig.~\ref{2diags}), defining a domain of stability of a
\textit{RK phase}, which has to be interpreted more appropriately as a domain
where the variational approach fails; (\textit{ii}) in the 
$t_{2}-W$ model when $\theta$ is close to $-\pi/2$, the failure of this
variational approach is expected since the exact ground state consists of
Heisenberg chains (weakly dimerized for finite $t_{2}$), with spin-spin
correlations along chains slowly decaying and the trial wave functions
describing isolated resonating plaquettes are no more valid there. For that
reason, the position of the \textit{MCPC-1/RSPC} transition, predicted in this
approach at $W/t_{2}=-4$ (hence $\theta=-1.326(1)$ relatively close to
$-\pi/2$) can differ appreciably from the real position of this transition. To
determine the latter, one has to treat the model exactly, taking all allowed
configurations into account (and not only those characteristic of the
\textit{plaquette} ordering) by methods such as exact diagonalisation.

\section{\label{exd} Analysis of the low-energy spectrum and identification of
  phases} 

In the limit $t_{2} \ll |W|$ of the $t_{2}-W$ model, where the variational
approach of Section \ref{vari} is least reliable, the kinetic term $H_{K}$ can
be described as a perturbation, while the unperturbed Hamiltonian $H_{W}$ has
for ground state a product of Heisenberg chains. The effect of $H_{K}$ can be
described by an effective coupling $K_{\rm{eff}}$ ~ $W|t_{2}/W|^4$, associated
with processes of order $4$ out of the Heisenberg ground state, where 2
singlets belonging to 2 neighbouring chains and opposite to each other are
flipped and then flipped back; the interaction resulting from this process is
attractive due to the $H_{W}$ term on the interchain plaquette(s), making this
process the most important one in perturbation in $t_{2}/W$ (among those having
an influence on the type of dimerization). Although we did not determine
analytically the sign of this effective coupling for $0 < t_{2} \ll |W|$, by
analyzing the energy splitting of the the analogs on the checkerboard of the
$(A,\pi)$ and $(B,\pi)$ excitations (of the 2-chains $J-K$ model) we determine
the sign of $K_{\rm{eff}}$, i.e. the type of dimerization occurring in the
system. 

\subsection{\label{lowt} Weak-coupling regime: low-energy spectrum and quantum
  numbers} 

\begin{figure}
\begin{center}
\includegraphics[width=8cm]{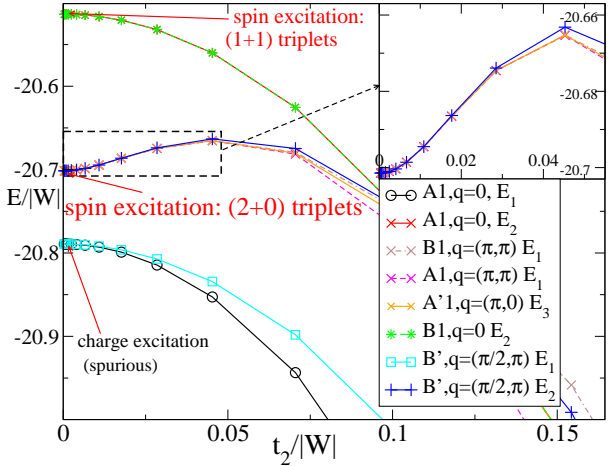}
\caption{\label{lowt2} (Color online)
Energies of lowest excited states (in the $S=0$ sector) of symmetries
$A1/B1, \textbf{q}=(0,0)/(\pi,\pi)$, $(A'1,(\pi,0))$, $(B',(\pi/2,\pi))$, at $t_{2}
\ll |W|$, for a cluster of $N=64$ sites. Inset: splitting between a 2-triplet 
$(B',(\pi/2,\pi))$ state (plus symbol) and other 2-triplet states (X symbols).}
\end{center}
\end{figure}

Let us consider first the $t_{2}=0$ limit: here the low-energy spectrum has a
simple structure, i.e. lowest states are composed of lowest states of
Heisenberg chain of the corresponding length (3 chains of 
$L=6$ for the $N=72$ cluster, 2 chains of $L=4.k$ for the $N=32.k$ cluster
$(k=1,2)$. The ground state is found in sectors $(A1,(0,0))$, $(B1,(0,0))$,
$(A'1,(\pi,0))$ and $(A'1,(0,\pi))$ (the 4-fold degeneracy accounts for the 4
ways of regularly accommodating 2 Heisenberg chains on the checkerboard
cluster). At $t_{2}=0$, and by extension in the weak-coupling regime $|t_{2}/W|
\ll 1$, the first excited states are, either on the cluster $N=64$ (see
Fig.~\ref{lowt2}) or $N=72$: {\it (i)} a state in the $S=1$ sector (degenerate
between various quantum numbers) corresponding to a 1-triplet excitation on one
chain and the Heisenberg ground state of the other(s) chain(s); {\it (ii)} a
state in the sector $S=0$, with all quantum numbers listed in the
table~\ref{Table}, corresponding to a charge excitation breaking of a Heisenberg 
chain into an isolated singlet and an open
$(L-2)$ chain); {\it (iii)} a state in the sector $S=0$ corresponding to 2
triplet excitation on the same Heisenberg chain, the other chains being in 
their ground states. On the $N=64$ cluster this state is found with the 
quantum numbers $(A1,(\pi,\pi))$, $(B1,(\pi,\pi))$, $(A'1,(\pi,0))$,
$(B',(\pi/2,\pi))$ (and those related by symmetry); {\it (iv)} slightly above
the latter, a state in the sector $S=0$ corresponding to two 1-triplet
excitations on distinct chains.

Although the lowest excitation in the $S=0$ sector is, on clusters considered, 
the charge excitation, as far as these states can be labeled as \textit{charge}
and 
\textit{2-triplet}-excitations we rather focus on 2-triplet excitations (states
	   {\it (iii)} and {\it (iv)}), since their excitation energy is a
	   finite-size effect, being proportional to $|W|/L$ which vanishes in the
	   thermodynamic limit, while state {\it (ii)} has an excitation energy of
	   order $|W|$ even for $L \rightarrow \infty$. Among 2-triplet
	   excitations, state  {\it (iii)} (2 triplets on the same chain) is of
	   greater interest: not only does it have a slightly lower energy than
	   state {\it (iv)}, but a reasoning based on the model situation of 2
	   Heisenberg chains coupled by a 4-spin coupling term (see Sec.~\ref{2LH})
	   shows that this state gives information about the type of dimerization
	   favoured by the interchain coupling (or $t_{2}$ here).

To analyse the influence of a weak kinetic coupling $t_{2} \ll |W|$ on the
system and determine, in the light of previously discussed features of the
$J-K$ model of coupled Heisenberg chains, we focus on states with 2 triplet
excitations on the same chain. The states corresponding to this excitation with
different quantum numbers split when $t_{2}$ increases; at lowest order in
$t_{2}/|W|$ the splitting occurs between the state in $(B',(\pi/2,\pi))$ (2D
analog of the $(B,\pi)$ 2-triplet state in 2 coupled Heisenberg chains,
hereafter \textit{B state}) and states with other quantum numbers (analogs of
the $(A,\pi)$ 2-triplet excitation for 2 coupled Heisenberg chains; hereafter
\textit{A states}). In this splitting the \textit{B state} (associated with a
dimerization in antiphase) has higher energy than the \textit{A states} 
(associated with a dimerization in phase); moreover, we have checked that the
energy difference between the \textit{B state} and the \textit{A states} scales
as $W(t_{2}/W)^4$, as expected from the previously discussed comparison with
the $J-K$ model of 2 Heisenberg chains (where the corresponding splitting
occurs linearly in $K/J$). Hence we conclude that on the checkerboard, for
small but finite $t_{2}/|W|$ the Heisenberg chains dimerize in phase and form a
\textit{MCPC-1} phase, which is consistent with results of Section \ref{vari}. 

While in the limit of small $t_{2}/W$ the physics of the model is quasi-1D,
making finite-size effects very important on the checkerboard clusters analyzed
(especially at $t_{2}=0$, Heisenberg chains being critical), for larger
couplings these finite-size effects become less relevant, as soon as the
clusters can accommodate the various ordered phases. Hence in the latter case
and with the cluster sizes available, it is reasonable to analyze the $t_{2}-W$
model non-perturbatively, and without making reference to an effective model
(such as that of coupled Heisenberg chains).

\subsection{Non-perturbative analysis for intermediate $W/t_{2}$:
  \textit{MCPC-1} or \textit{RSPC} phase~?} 

\begin{figure}[!h]
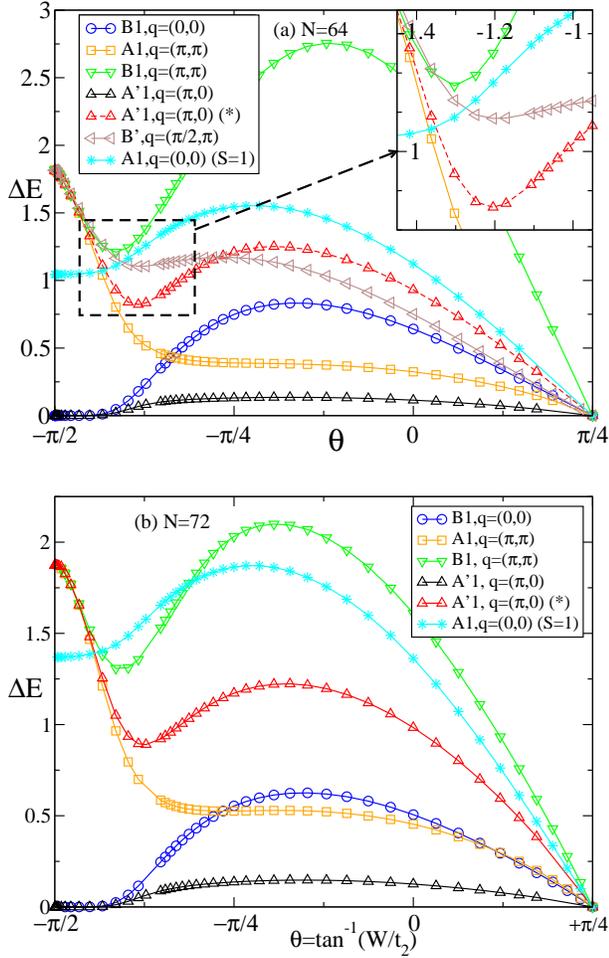

\begin{center}
\includegraphics[width=8cm]{DEvarW32_52.eps}\\
\parskip 15pt
\includegraphics[width=8cm]{DE_Wvar66.eps}
\caption{\label{Ext2pW} (Color online) Excitation energies $\Delta E=E_{0/1}-E_{GS}$ ($E_{1}$
  if labeled by $(*)$) in units of $\sqrt{t_{2}^2+W^2}$ for different quantum
  numbers, in the singlet \cite{S01} magnetic sector (open symbols) as a
  function of $\theta=\arctan(W/t_{2})$, on the $N=64$(a) and $N=72$(b)
  checkerboard clusters. The lowest triplet excitation is also shown (star
  symbols).} 
\end{center}
\end{figure}

Away from the $t_{2} \ll |W|$ limit, one cannot simply identify each of the
first excitations as \textit{2-triplet} or \textit{charge} excitations, but by
using symmetries of the model one can characterize them by their quantum
numbers. As seen in Sec.~\ref{diag}, for each of the candidate phases, in the
thermodynamic limit, we know the quantum numbers associated with wave functions
of the (4- or 8- degenerate) ground state. Hence the relative order (in energy)
of first excitations with those quantum numbers, for large enough systems,
should be characteristic of the symmetry of the ground state, and thus of the
phase in question. From Sec.~\ref{lowt}, the \textit{MCPC-1} phase is expected
to extend over a finite range of $W/t_{2}$, going either to the RK point or to
a non-trivial point of transition towards a RSPC phase (in analogy with the QDM
on the square lattice). In Fig.~\ref{Ext2pW} we plot the lowest excitation
energies (in units of $\sqrt{t_{2}^2+W^2}$) corresponding to quantum numbers
listed in Table~\ref{Table}, as a function of the parameter
$\theta=\arctan(W/t_{2})$ (N.B. This is equivalent to considering the
Hamiltonian $H=\sin(\theta) [H_{W}(W=1)] + \cos(\theta) [H_{K}(t_{2}=1)]$ ). 

The most striking feature of these graphs concerns the states with quantum
  numbers $(A1,(\pi,\pi))$ and $(B1,(0,0))$. The excitation
  energy (in units of $\sqrt{t_{2}^2+W^2}$) of the former, being that of the
  charge excitation in the $\theta \rightarrow -\pi/2$ limit, collapses when
  $\theta $ increases. In comparison, the $(B1,(0,0))$ state, degenerate with
  the ground state at $t_{2}=0$ (uncoupled Heisenberg chains) becomes separated
  energetically from the ground state when $t_{2}$ becomes non negligible (the
  vanishing of both excitation energies when $W/t_{2}$ gets close to $1$ is due
  to the degeneracy at the RK point). On the $N=64$ cluster, for a wide range
  of $\theta$, the $A1,(\pi,\pi)$ state has an energy significantly
  lower than the $(B1,(0,0))$ state and close to that of lowest $(A1,(0,0))$
  and $(A'1,(\pi,0))$ states (This is less obvious with data from the $N=72$
  cluster, where the effective length of chains $L=6$ results in stronger
  finite-size effects). The eigenvalue crossing between $(A1,(\pi,\pi))$ and
  $(B1,(0,0))$, similar to that observed for $n=1/2$, indicates a
  breaking/restoration of rotational symmetry, signaling a transition between
  the \textit{MCPC-1} phase (expected in the thermodynamic limit for $t_{2} \ll
  |W|$) and the RSPC for which the symmetrized wave functions of the ground
  state have quantum numbers $(A1,(0,0))$, $(A'1,(\pi,0))$, $(A'1,(0,\pi))$ and
  $(B1,(\pi,\pi))$. As the data of the $N=72$ cluster are less clear, the
  present analysis should be extended to larger clusters, which is unrealistic
  with the current most advanced computational resources \cite{N80}
  unless we find other tools to be more conclusive about the existence and
  position of a \textit{MCPC-1/RSPC} phase transition. 

\section{\label{corr} Plaquette correlations on the checkerboard}

Since the finite-size effects encountered for clusters of sizes $N \le 72$,
particularly in the weak-coupling regime and at the supposed
\textit{MCPC-1/RSPC} transition, make it difficult to identify clearly this
transition by analysing the low-energy spectrum only, one needs complementary
information about the nature of the ground state. The ED numerical technique
employed in this study allows also to compute expectation values of observable
and their associated correlations. Considering the structure of the different
ordered phases expected in this model (both the \textit{MCPC-1} phase and the
\textit{RSPC} are composed of singlets localized on uncrossed plaquettes of the
checkerboard), it is preferable to consider an observable defined on an
uncrossed plaquette (and then symmetrized) rather than on a single site.
Hence in this section we discuss results about two types of plaquette
correlation functions in the ground state of the $t_{2}-W$ model, computed on
clusters $N= 32,64,72$. 

The first type of plaquette operators for which we compute correlations have
$B1$ point group symmetry, and are related to the \textit{flippability}
of a given plaquette located at $\textbf{r}$ (i.e. of an uncrossed plaquette
of the checkerboard): 
\begin{eqnarray}
P_{-}(\textbf{r}) = (n_{1,h}n_{2,h} - n_{1,v}n_{2,v}) \delta_{S_{1,z}+S_{2,z}} 
\end{eqnarray}
The correlations of $P_{-}$ operators are computed from the ground state wave
function $|\Psi_{GS}\rangle$ (first determined as the ground state in the 
$A1,\textbf{q}=(0,0)$ symmetry sector and then expanded on all configurations). The
average value of $P_{-}$ on the ground state being zero by symmetry (for any
plaquette), the correlation function is defined as: 
\begin{eqnarray*}
C_{-}(R)=n_{R}.\langle \Psi_{GS}|P_{-}(0)P_{-}(\textbf{r}) |\Psi_{GS}\rangle
\end{eqnarray*}
where $P_{-}(0)$ is computed on a reference plaquette, $P_{-}(\textbf{r})$ on a
plaquette at distance $R$ from the reference one and $n_{R}$ is the number of
plaquettes at distance $R$ from the reference. A non-zero value of $C_{-}(R)$
in the thermodynamic limit and at large distances corresponds to a phase
breaking the $\pi/2$-rotational symmetry (hence in this context to the
\textit{MCPC-1} phase). 

\begin{figure}
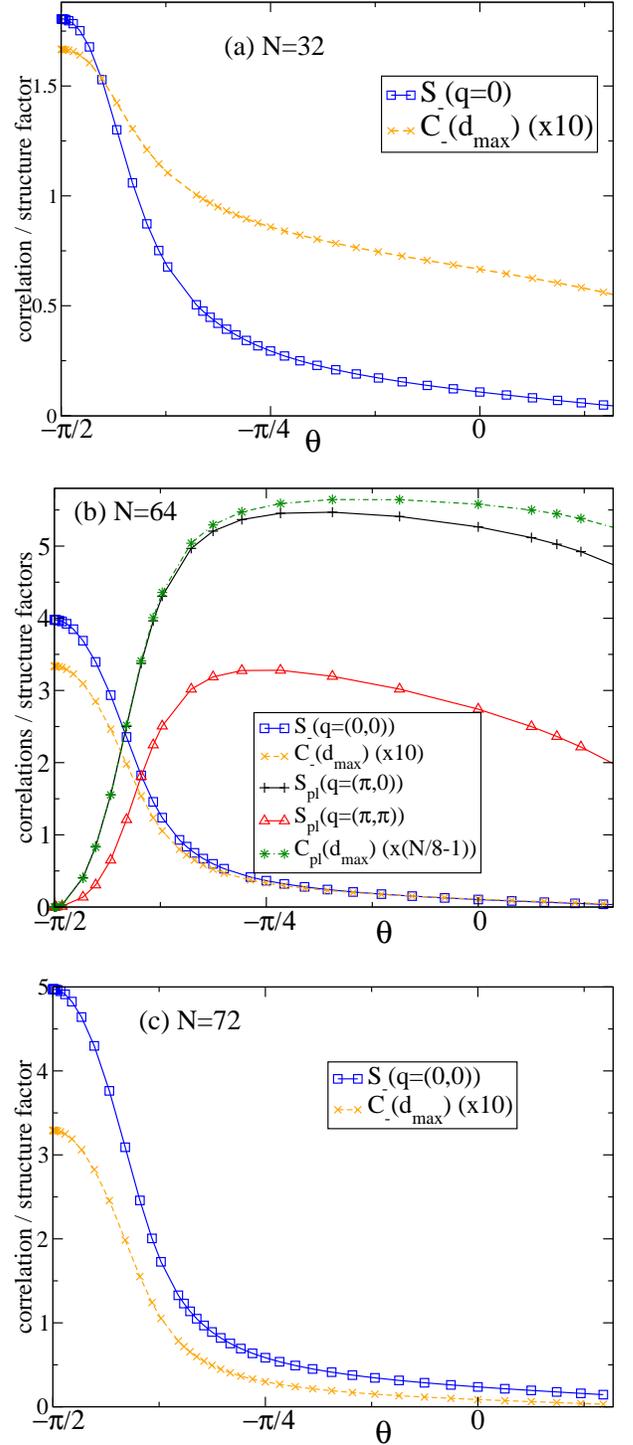

\begin{center}
\includegraphics[width=8cm]{corrAB1N16_192.eps}\\
\vspace{0.4cm}
\includegraphics[width=8cm]{corrAB1N32_203.eps}\\
\vspace{0.4cm}
\includegraphics[width=8cm]{corrAB1N36_193.eps}
\caption{\label{cor1} (Color online) Correlations in real space ($C_{-}$) at distance
  $d_{max}$ (dashed and dotted curves), and in reciprocal space ($S_{-}$) at
  $\textbf{q}=0$ (square symbols) of $B1$ plaquette operators, on clusters of
  size $N=32$ (a), $N=64$ (b) and $N=72$ (c), as a function of
  $\theta=\arctan(W/t_{2})$. On graph (b) are also shown, for the $N=64$
  cluster, plaquette-exchange correlations $C_{pl}$ at distance $d_{max}$ and
  the corresponding structure factor $S_{pl}$ at $\textbf{q}=(\pi,0)$ and
  $\textbf{q}=(\pi,\pi)$.} 
\end{center}
\end{figure}

In Fig.~\ref{cor1} are shown, for each of the cluster sizes considered,
correlations $C_{\alpha}(R=d_{max})$ at the maximal distance between plaquettes
equivalently occupied in a \textit{plaquette} phase ($d_{max}=2\sqrt{2}$ for
$N=32, 72$ and $d_{max}=4$ for $N=64$). We also plot Fourier transforms $S_{-}$
of $C_{-}$ correlations, for wave vectors $\textbf{q}=(0,0)$ and
$\textbf{q}=(\pi,\pi)$. These Fourier transforms are computed with a truncation
at short distances, i.e. only correlations for distances $r \ge 2$ are taken
into account (distances $R=0,1$ are discarded since the plaquette operators on
neighbouring plaquettes share at least one site; at distance $R=\sqrt{2}$ the
simultaneous double occupancy of both plaquettes is forbidden by the
\textit{dimer constraint}).

The correlations of the $B1$ plaquette operator ($P_{-}$) vary significantly
with $\theta$ between the Heisenberg- and the RK- limits, and give important
information about the evolution of the ground state. Correlations of $P_{-}$
decrease strongly for $\theta \le -3\pi/8$ (which means $W/t_{2} \le W_{c} \sim
-2.4$), both at the largest distance $d_{max}$ between equivalent plaquettes,
and in Fourier space at wave vector $\textbf{q}=(0,0)$ -in the weak-coupling
regime (for $\theta \le \theta_{c} \sim -1.2(1)$) $S_{-}(\textbf{q}=(0,0))$ is
well approximated by a Gaussian function of $\theta + \pi/2$; for $\theta \ge
\theta_{c}$, the ratio $S_{-}(\theta)/S_{-}(-\pi/2)$ becomes smaller when
$d_{max}$ increases, so one can expect $B1$ plaquette correlations to vanish in
the thermodynamic limit. This indicates that the rotational symmetry of the
lattice, broken in the weak-coupling regime, is restored for $\theta \ge \theta_{c}$, 
which is a signature of a transition to a \textit{RSPC} phase.

We also computed (off-diagonal) plaquette-exchange correlations,
i.e. correlations of the kinetic operator 
$$P_{\square}(r)=(b^{\dagger}_{i\uparrow}b^{\dagger}_{j\downarrow}+
b^{\dagger}_{i\downarrow}b^{\dagger}_{j\uparrow})\nn\\  
(b_{k\uparrow}b_{l\downarrow}+b_{k\downarrow}b_{l\uparrow}) +c.c.$$
(where sites $i,j,k,l$ are those around the void plaquette at position
$r$). The connected correlations $C_{pl}(r)=\langle
P_{\square}(r)P_{\square}(0)\rangle -\langle P_{\square}(0) \rangle^2$ are
vanishing in the $\theta \rightarrow -\pi/2$ limit, where charge moves away
from Heisenberg chains are energetically forbidden; at the contrary, in a
\textit{RSPC}, they are expected to be important between \textit{resonating
  plaquettes} (at relative position $(2.p,2.q) - p,q \in \mathbb{Z}$ from each
other) and significantly smaller otherwise. Consequently we focus specifically
on plaquette-exchange structure factors $S_{pl}(\textbf{q})$ (Fourier
transforms of the corresponding correlations) at $\textbf{q}=(\pi,0)$,
$\textbf{q}=(0,\pi)$ and $\textbf{q}=(\pi,\pi)$. They are shown for the $N=64$
cluster, along with the correlation in real space at maximal distance
$C_{pl}(d_{max})$, on Fig.~\ref{cor1} (graph (b)).  

These correlations increase significantly with $\theta$ in the weak-coupling
regime, indicating the appearance of resonating plaquettes characterizing
the \textit{MCPC-1} and \textit{RSPC} phases. We have checked (not shown) that
the $(\pi,0)$ structure factor is very well approximated by its contribution
from correlations between resonating plaquettes only, which indicates that
the picture of a \textit{plaquette pattern} describes well the ground state
wave function at $\theta \ge \theta_{c}$, where these correlations are
important. The smaller value of the structure factor at $\textbf{q}=(\pi,\pi)$
compared to that at $\textbf{q}=(\pi,0)$ originates from non-negligible
correlations at short distances such as $d=\sqrt{5}$. The comparison between
the correlation $C_{pl}(d_{max})$ rescaled by a factor $N/8-1$ (accounting for
the number of plaquettes included in the Fourier sum defining $S_{pl}$ and that
should be occupied and resonating in a \textit{MCPC-1} or \textit{RSPC} phase,
and the structure factor $S_{pl}(\pi,0)$ itself, is eloquent:
it indicates that for $\theta \le \theta_{c}$, almost all contributions to the
Fourier transform come from resonating  plaquettes, and the correlations
between these plaquettes are long ranged. For larger values 
of $\theta$ ($ \ge \theta_{c}$) the (negative) contributions from other
plaquettes become non-negligible; but structure factors at both wave vectors
remain significant, and indicate the robustness of the ordering on these
plaquettes. Although plaquette-exchange correlations alone 
are not sufficient to distinguish the \textit{RSPC} from the \textit{MCPC-1}
phase, their change of behaviour around $\theta = \theta_{c}$, associated with
features of the low-energy spectrum and $B1$ (diagonal) plaquette correlations,
indicates the presence of a \textit{MCPC-1/RSPC} phase transition.

\section{Conclusion and perspectives}

In summary, we have considered $S=1/2$ fermions on the checkerboard lattice,
with an extended Hubbard model in a limit of infinite on-site  and strong
nearest-neighbour Coulomb repulsions, at the specific fractional $1/8$
filling. In this limit, constraints characteristic of the square lattice dimer
model naturally emerge, with the links on the square lattice corresponding to
the sites of the checkerboard. Moreover, the lowest-order kinetic process
allowed in perturbation (from the infinite repulsion limit) flips 2 particles
around an uncrossed plaquette, recalling the kinetic term of the Quantum Dimer
Model, and the analogy is reinforced by considering an extra term similar to
the potential term of the QDM. However, here the spin degrees of freedom of
particles play an essential role, since kinetic processes act only on singlet
states on a plaquette. The model we have considered makes a continuous
connection between the physics of critical Heisenberg chains (occurring in
place of the \textit{columnar} limit of the QDM) and a \textit{RK-like}
critical point of the present model. 

Starting from a situation where the Heisenberg chains of the first case are
weakly coupled, we have identified the leading order - in perturbation in 
$t_{2}/W$ - term coupling neighbouring chains, and characterized it as a 
relevant perturbation for the Heisenberg chains, driving their in-phase 
dimerization in the thermodynamic limit as soon as the coupling is finite.
The corresponding phase (\textit{Mixed Columnar-Plaquette Crystal -1}) 
distinguishes itself from other candidate phases (\textit{RSPC} and \textit{columnar})
of the model by a lower symmetry, and its extent in parameter space is determined by 
a detailed analysis of the low-energy spectrum obtained by exact diagonalization, 
taking lattice- and time reversal- symmetries into account. This analysis indicates
that, when going towards the {RK-like} point, the  
\textit{MCPC-1} phase persists up to a transition into a $\pi/2$-rotational
invariant \textit{Resonating Singlet Pair Crystal}, the analogue of the
plaquette phase of the square lattice QDM; this transition is confirmed by the
computation of various types of correlations between uncrossed plaquettes of
the checkerboard, characterizing the plaquette ordering, and the symmetries of
the ground state. The qualitative features of this phase diagram are also found
by a variational approach, also indicating that the transition between the
\textit{MCPC-1} phase and \textit{RSPC} should be of second order. In
particular, the system described by the extended-Hubbard model with strong
repulsions considered first appears to be in a \textit{Resonating Singlet Pair
  Crystal}, similarly to the corresponding model at quarter filling~\cite{PPS}.

\begin{figure}
\begin{center}
\includegraphics[width=6cm]{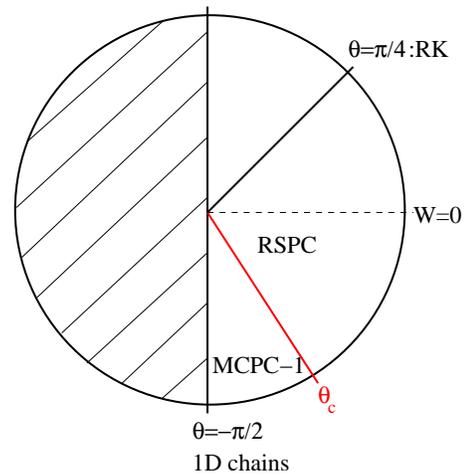}
\caption{\label{camtw} Phase diagram of the $t_{2}-W$ model of $S=1/2$ fermions
  at $n=1/4$ (in the range of parameter $-\pi/2 \le \theta \le \pi/4$ where 
$\theta = -\arctan(W/t_{2})$). The \textit{MCPC-1/RSPC} transition occurs at 
$\theta_{c} \sim -3 \pi/8 \pm 0.2$ according to exact diagonalization results 
(a value $\theta_{c}=\arctan(-4)$ is estimated variationally).} 
\end{center}
\end{figure}
An open question is to know what happens at small but finite doping from the
$1/8$ filled case: a possibility is that the system remains in a crystalline
phase confining the doping particles (either holes or electrons depending on
the type of doping); alternatively, the \textit{RSPC} could give way to a
phase with either deconfined doping particles or the formation of bound Cooper
pairs (as it happens in systems at small doping from $1/4$ filling and $1/2$
filling~\cite{check12}) that would be an indication of a superconducting or
supersolid phase. 

FT and DP thank S. Capponi for discussions on the two-coupled chain problem , the 
French Research Council (ANR) for financial support, and IDRIS (Orsay, France) for 
computation time. Likewise, RM thanks Oleg Starykh for explanations of his related 
work. RM and DP thank the Kavli Institute for Theoretical Physics, where this 
collaboration was initiated, for hospitality.


\begin{thebibliography}{99}
\bibitem{RKvar} Close to $W=t_{2}$ and in the thermodynamic limit, the RK wave
  function has an energy of $2(W-t_{2})n_{fl}$ per uncrossed plaquette (so
  twice as much per particle), where $n_{fl}$, the probability for a given
  plaquette to be flippable, equals $1/4$. Similarly, in the bosonic case, the
  flippability is $n_{fl}=1/4$ and the energy of the RK state equals $(V-t)/2$
  per particle.
\bibitem{S01} Using the spin inversion symmetry allows one to compute
  eigenstates and eigenenergies in sectors of total spin either odd or even,
  but not explicitly in the $S=0$ or $S=1$ sector. However it is known that
  states of lowest energy are of total spin $S=0$ or $S=1$.
\bibitem{N80} Two clusters of size $N=80$ would be tractable by exact
  diagonalization, but there either the $\pi/2$ rotation symmetry or the
  reflexion symmetry would be lost (in comparison to the $N=64$ cluster for
  instance) which would be problematic for determining the quantum numbers
  characterizing the phases of the model. The \textit{next} cluster available
  (by increasing $N$) and possessing these symmetries would be the $N=128$
  cluster, far beyond the limits of existing numerical resources for Exact
  Diagonalizations.
\end{thebibliography}
\end{document}